# Exploring Trust and Risk during Online Bartering Interactions

KALYANI LAKKANIGE, The Pennsylvania State University, USA
LAMAR COOLEY-RUSS, The Pennsylvania State University, USA
ALAN R. WAGNER, The Pennsylvania State University, USA
SARAH RAJTMAJER, The Pennsylvania State University, USA

This paper investigates how risk influences the way people barter. We used Minecraft to create an experimental environment in which people bartered to earn a monetary bonus. Our findings reveal that subjects exhibit risk-aversion to competitive bartering environments and deliberate over their trades longer when compared to cooperative environments. These initial experiments lay groundwork for development of agents capable of strategically trading with human counterparts in different environments.

## 1 INTRODUCTION

The act of bartering requires both parties to take some risks. For instance, each party is likely uncertain about the quality of the item that they will receive [1]. They may also not have an accurate understanding of the value the object they wish to trade. Moreover, in situations involving bartering, one trader may take advantage of the other party's need for an item. That is, when a trader is most in need of an item, and hence the most vulnerable, an unscrupulous trading partner might take advantage and rapidly increase the price of the sought-after item.

The risks associated with bartering can be mitigated, in part, by building relationships with potential trade partners. These relationships use the history of trade interactions with a partner as a basis for accurately predicting the expected outcome of some future trade. Over time trade partners establish trust in each other because they develop accurate shared expectations which reduce the risks associated with trading with that person [13].

Our work seeks to understand how people barter and how trusting relationships for trade are established and maintained. There has been some research formalizing barter as a means of trade [4, 9, 12]. Shapley and Scarf, for instance, developed a game theoretic framework for modeling barter as a game [10]. Their foundational work has been further developed through increasingly complex scenarios to emulate real world trading systems [2, 5, 7].

The exploratory work presented here examines bartering behavior under competitive, cooperative, and baseline (neither competitive or cooperative) conditions. We explore the hypothesis that, when bartering in a competitive

Authors' addresses: Kalyani Lakkanige, kxl5656@psu.edu, The Pennsylvania State University, University Park, State College, Pennsylvania, USA, 16801; Lamar Cooley-Russ, The Pennsylvania State University, University Park, State College, USA, lbc5186@psu.edu; Alan R. Wagner, The Pennsylvania State University, State College, Pennsylvania, USA, 16801, azw78@psu.edu; Sarah Rajtmajer, The Pennsylvania State University, State College, Pennsylvania, USA, 16801, smr48@psu.edu.

situation, traders will tend to be more secretive, or conservative with information-sharing. Our experiments take place within an online gaming platform, so secretive behavior is operationalized as direct (vs. global) messaging. Direct messages limit the information given to potential competitors and, from a game theory perspective, serve a strategy for limiting risk and exposure [6]. We further hypothesize that, because competitive games present greater risk, each potential trade will require more deliberation on the part of the trader. We conjecture that the presence of risk motivates traders to closely consider the costs and benefits of accepting a trade and thus acts as an extra form of cognitive workload. This additional workload increases the amount of time taken before they accept. Our findings support these hypotheses and highlight the significance of environmental influence on bartering behavior.

## 2 METHODS

We have created a Minecraft plugin called BarterPlus [3, 11] to run experiments, gather data and simulate a barter market for competitive, cooperative, and baseline scenarios [2, 8]. We developed GUI tools which allow players to observe and decide on items to trade [3]. Each player was assigned a profession and given a randomized list of items before the game began. Once the game started, players assumed the role of traders to discuss and exchange items with potential trade partners. We were able to monitor their interactions and logged every aspect of trades. Each barter session ran for thirty minutes and players were expected to exchange items according to their role to gain *points*, which are meta values attached to each item a profession requires. Rarity of items determined their value. A tier-1 item would value a point each and there were 20 for each profession. Similarly, there were 10 tier-2 items valued at three points and only 3 tier-3 items for each profession. To incentivize players to continue trading until the end of game, they were not told how many of each items were available and were not able to create items. The players were asked to use the chat built into the environment to *talk* with each other. Traders could discuss their trades in two ways. They could either communicate in the global chat or directly message a potential trade partner.

The data recorded from these experiments include a JSON file, log file and participant feedback form. Our plugin kept record of the status and events pertaining to a given game in a JSON file. This data included each participant's username, their profession, the items they were required to obtain, their final score at the end of the game and information about each *TradeRequest. TradeRequests* are events that occur when a player attempted to initiate a trade with a potential trade partner. Information in a TradeRequest include the start time, end time, initiating trader (known as the requester), the recipient trader (known as the requested), the involved traders' scores after each trade and the items each trader offered in an exchange. TradeRequests can have one of three different status: *accepted*; *cancelled*; or, *denied*. An accepted request is one wherein both players agreed to exchange items denoted in the trade placed in the trade menu. A cancelled trade is one wherein the requester withdrew their request. A denied trade is one wherein the requested rejected the request. The log file includes the raw output from the server console which includes timestamped logs of all global chat communications, all direct messages, and the details of each trade as they occurred.

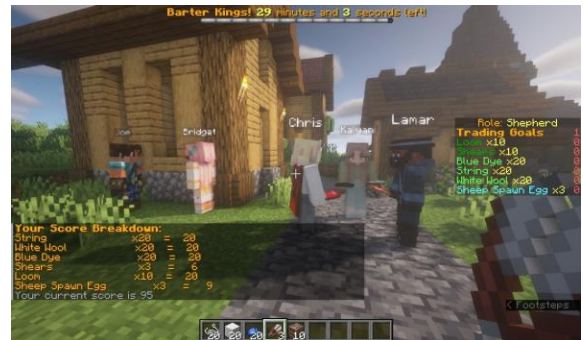

Fig. 1. Screenshot from a game. It shows traders gathered near each other to discuss trades. It also shows information available for a player such as, the time left for the game to end, their profession, items they must acquire, value of each item, items in their inventory, current score and a chat bar.

Table 1. Overview of experiments

| Game type | Competitive | Baselines | Cooperative |
| --- | --- | --- | --- |
| No. of games | 7 | 6 | 7 |
| No. of players | 39 | 33 | 37 |
| Players who earned bonus(%) | 17.95 | 78.79 | 86.49 |
| No. of trades initiated | 544 | 485 | 541 |
| No. of accepted trades | 215 | 188 | 236 |
| No. of total chats | 2058 | 1841 | 2152 |
| No. of direct messages | 421 | 184 | 192 |
| $25^{th}$ percentile of trade acceptance time(s) | 9.53 | 9.68 | 9.37 |
| $75^{th}$ percentile of trade acceptance time(s) | 25.32 | 19.46 | 19.57 |

A total of 109 subjects participated in our IRB-approved study. We conducted twenty-two runs, each involving 5-7 players. Data from 2 runs was discarded because of too few participants and a technical error. Research participants were recruited using flyers posted on the university campus and posts on university specific subreddits and Discord servers. The eligibility criteria to participate included being at least 18 years old, having Minecraft Java edition installed on a computer to enable remote connection to our server, and having prior experience playing Minecraft. Participants were paid $10 for the hour-long study and could earn an additional $5 bonus if the value of the items they obtained exceeded a prescribed threshold, disclosed to them at the start of the experiment. Each experiment was one of three types: competitive; cooperative; or, baseline. In the competitive condition, only one player, i.e., the player with the most points, earned the bonus. In the cooperative condition, all players obtained the bonus if every player collected a sufficient number of the items to meet the prescribed threshold. In the baseline condition, every player obtaining number of points at or above the threshold was able to earn the bonus.

Players joined the study using a privately shared Zoom link. This ensured clear communication of the instructions and opportunity for clarifications, if necessary. At any point in the session, players could check their score. At the end of the 30 minutes, the server closed forcing everyone out of the study. Participants would then be prompted to provide feedback about the study and any strategies they used during the game.

## 3 RESULTS

A total of 109 players ($N = 109$) participated in the experiment. Table 1 presents statistics from the study. With respect to the hypotheses, we note that subjects used global messaging significantly more ($p = 0.02$) in the cooperative condition than in the competitive or baseline, as expected.

Likewise, results show significantly more direct messaging in the competitive condition ($\mu = 23.52\%, \sigma = 31.61\%$) where the risks of not earning the bonus was higher than in the cooperative condition ($\mu = 10.43\%, \sigma = 13.11\%$). This finding is supported by a two-tailed t-test ($p = 0.02$). Fig. 2 presents a box plot of all players' percentage use of direct messages for each condition. Some of the traders from the three scenarios do not use direct messages at all. The maxima of the competitive game indicate a minority of players opting for global message.

Our results also suggest that traders experience greater workload in the competitive situation. In support of our hypotheses, we find that the trade acceptance time is longer in the competitive condition ($\mu = 27.28s, \sigma = 68.09s$) than in the cooperative condition ($\mu = 18.54s$, $\sigma = 42s$). This is supported by a two-tailed t-test ($p = 0.02$). Fig. 3 presents time to accept trades in the three game types. Small differences in lower values indicate traders take similar minimum time to accept a trade. The upper value of acceptance time is approximately same for baseline and cooperative scenarios but traders in competitive scenario take significantly longer to accept a trade. Table 1 shows that similar behavior holds for the $25^{th}$ and $75^{th}$ percentiles, suggesting that competitive behavior is visibly different from other game types in the upper quartiles.

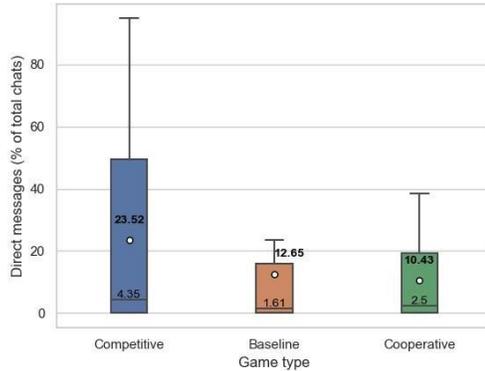
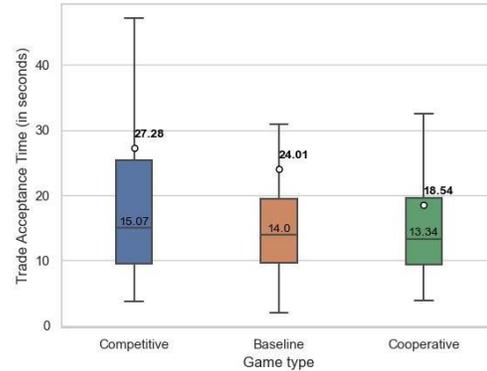

Fig. 2. Percent use of direct messages as mode of communication in competitive, baseline and cooperative trade scenarios. Includes mean percent use of direct messages (in bold) and median.

Fig. 3. Trade acceptance time in competitive, baseline and cooperative trade scenarios. Includes mean acceptance time (in bold) and median.

## 4 CONCLUSION

Our results show that traders' communication strategies are significantly influenced by the condition. In the competitive environment, traders use significantly more direct communications. Presumably they recognize that global messaging in this environment could provide valuable information to competitors and reduce their chance of obtaining the bonus. In contrast, in the cooperative condition, traders broadly share the content of their inventory and the list of items they need. This behavior reflects the fact that risks associated with sharing information are low in this condition.

Our data also indicates that traders spend significantly greater time thinking about trades before accepting them in the competitive condition than in the cooperative condition. Here again, risk likely influences traders to carefully consider the consequences of accepting a deal and whether better opportunities might be available in the future. In the cooperative condition, the risk associated with a bad trade is minimal. Hence, little thought is needed.

The data we are collecting is rich. Hence this work represent only a surface level preliminary analysis how risk and trust effect trading behavior in barter situations. Subsequent experiments will examine how trust evolves and attempt to create agents capable of strategically trading in different environments.

## ACKNOWLEDGMENTS

Portions of this material is based on work supported by the Air Force Office of Scientific Research (AFOSR) under award number FA9550-21-1-0197.